\def\ROSAT{{\it ROSAT\/\ }}
\def\ASCA{{\it ASCA\/\ }}

\def\Ic{IC~10\ }
\def\Icc{IC~10}

\def\ltsima{$\; \buildrel < \over \sim \;$}
\def\simlt{\lower.5ex\hbox{\ltsima}}
\def\gtsima{$\; \buildrel > \over \sim \;$}
\def\simgt{\lower.5ex\hbox{\gtsima}}

\documentstyle[psfig]{mn}

\title[\ROSAT HRI observations of \Icc, NGC~147 and NGC~185]
{\ROSAT HRI observations of the Local Group galaxies \Icc, NGC~147 and NGC~185}

\author[W.N. Brandt, M.J. Ward, A.C. Fabian \& P.W. Hodge]
{\parbox[]{6.5in}{W.N. Brandt,$^{1,2}$\thanks{Current address:
The Pennsylvania State University,
Department of Astronomy and Astrophysics,
525 Davey Lab,
University Park, Pennsylvania 16802, USA}
M.J. Ward,$^3$ 
A.C. Fabian$^2$ and P.W. Hodge$^4$}\\
\\
$^1$Harvard-Smithsonian Center for Astrophysics, 60 Garden Street, 
Cambridge, Massachusetts 02138, USA\\
$^2$Institute of Astronomy, Madingley Road, Cambridge CB3 0HA\\
$^3$X-ray Astronomy Group, Department of Physics \& Astronomy, 
University of Leicester, University Road, Leicester LE1 7RH\\
$^4$Department of Astronomy, Box 351580, University of Washington, 
Seattle, Washington 98195, USA\\
}

\begin{document}

\maketitle

\begin{abstract}  
We report on pointed X-ray observations of \Icc, NGC~147 and NGC~185 
made with the \ROSAT High Resolution Imager (HRI). These are three
Local Group galaxies that have never been 
previously studied in detail in the 
X-ray regime. \Ic is the closest starburst galaxy to 
our own Galaxy, 
and NGC~147 and NGC~185 are companions to M31. 
We have discovered 
a variable X-ray source coincident with
\Icc. The source is located near the centre of a large, non-thermal 
bubble of radio emission, and it is positionally 
coincident with an emission line star in \Ic which has been 
classified as a WN type Wolf-Rayet star. 
We demonstrate that a confusing foreground or background 
source is improbable. 
The X-ray source is probably an X-ray binary in \Icc, and 
it may be a Wolf-Rayet + black hole binary. 
The source has mean and maximum 0.1--2.5~keV isotropic 
luminosities of about 2 and 4 times $10^{38}$ erg s$^{-1}$.
We do not detect any sources 
in the central regions of NGC~147 or NGC~185.
We place upper limits on their central X-ray emission,
and we list all X-ray sources coincident with their
outer extents. We also present 
the first X-ray detections of the well-studied Algol-type binary 
TV Cas and the W UMa-type binary BH Cas, which were both serendipitously
observed during our \Ic pointing. 
\end{abstract}


\section{Introduction} 

\Ic is a small irregular galaxy in the Local Group 
(see Hodge 1994 for a brief history and references). 
It is rich in newly formed massive blue stars and is peppered
with over 140 H~{\sc ii} regions (Hodge \& Lee 1990). Recently,
Massey, Armandroff \& Conti (1992) and Massey \& Armandroff (1995) have 
identified 15 Wolf-Rayet stars in \Icc. Since Wolf-Rayet
stars are descended from only 
the most massive ($\simgt$30--40 M$_\odot$) stars,
this argues that \Ic has an unusually large unevolved massive (OB)
star population as well, with a {\it galaxy-wide\/} surface density of 
massive stars that is higher by a factor of 2 than any other
Local Group galaxy. The galaxy-wide massive star density of \Ic is 
comparable to that observed in isolated regions of recent star formation 
in other Local Group galaxies, in accord with earlier suggestions 
that \Ic is currently undergoing a starburst. 

\begin{figure*}
\centerline{\psfig{figure=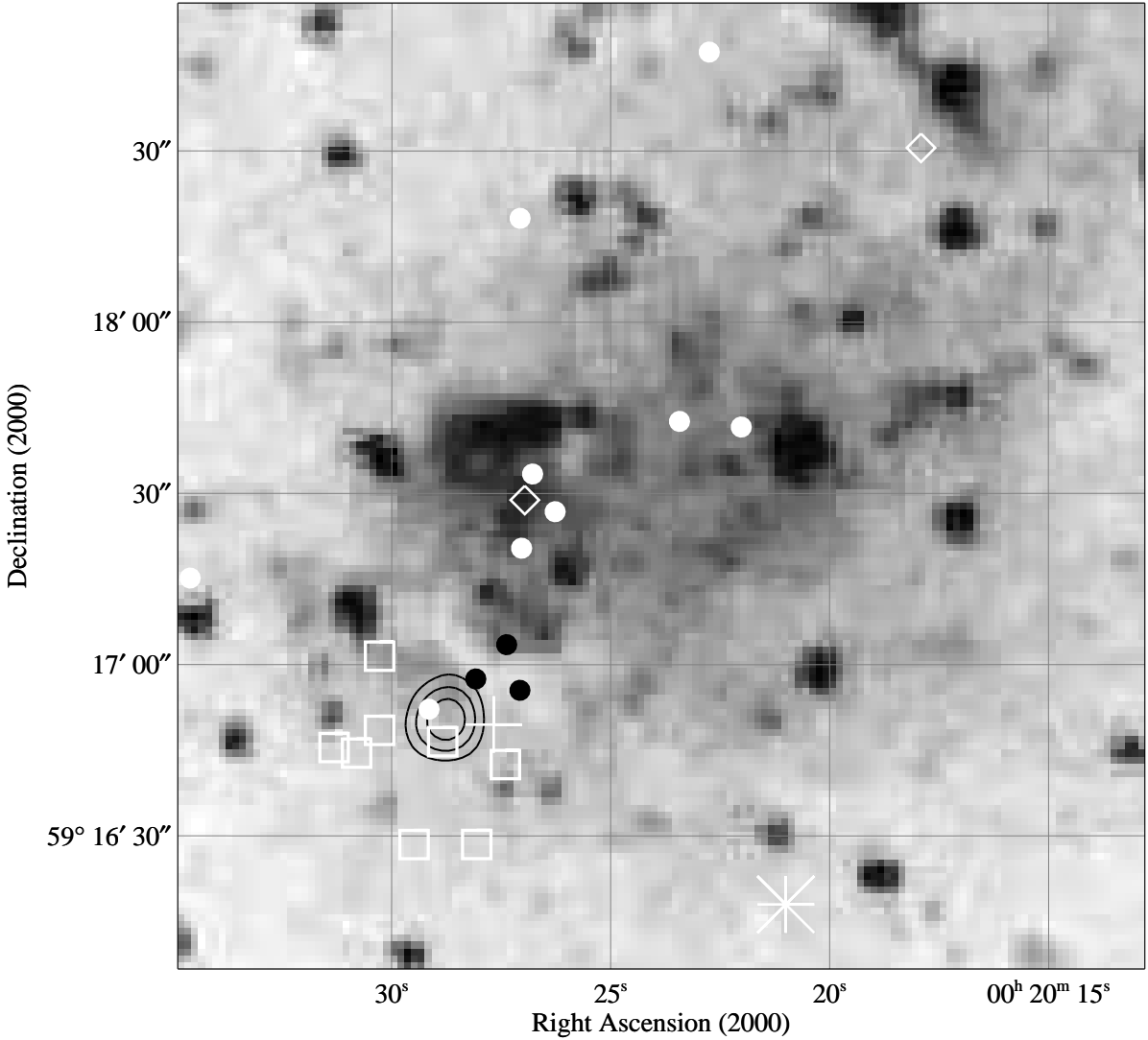,width=0.9\textwidth}}
\caption{Contours of the HRI image of \Ic overlaid on the image from 
the Palomar Schmidt $B_{\rm J}$ plate. Contours are in black and are at 
44.8, 67.1 and 89.5 per cent of the maximum pixel value (see the text for 
absolute normalization). White solid dots show the positions of probable 
Wolf-Rayet stars from Massey \& Armandroff (1995), and black solid dots
show the positions of molecular clouds from Wilson (1995). Diamonds show
the positions of maser complexes, and squares show the positions of SNR
candidates from table 3 of YS93. The cross shows the central position of  
H~{\sc ii} region \# 113 from Hodge \& Lee (1990) after correcting for
the offset given in Ohta et~al. (1992). The star shows the revised 
centroid position of the all-sky survey source, as described in 
the text (Th. Boller, private communication). Note 
that our HRI source is near the centre of the YS93 non-thermal 
radio superbubble (compare with Figure 3) and coincident 
with the claimed Wolf-Rayet star WR17.}
\end{figure*}

The radio continuum emission from \Ic is strong 
(e.g. Klein \& Gr\" ave 1986). Of particular
relevance to this work is the fact that 
Yang \& Skillman (1993; hereafter YS93) have 
discovered a non-thermal bubble of radio emission in \Icc. This 
bubble has a diameter of about 250 pc, 
and comparison with data at other wavelengths shows that 
it is associated with a region of star formation 
as well as the densest H~{\sc i} concentration in \Icc. 
YS93 argue that the bubble is a collection of supernova
remnants (i.e. a `superbubble'). The 
408~MHz radio luminosity of the bubble is
about 2.6 times that of N132D, one of the brightest 
supernova remnants (SNR) in the Large Magellanic Cloud.  
\Ic also has two H$_2$O maser complexes which include
a flaring megamaser as well as a maser with intraday 
variability (e.g. Argon et~al. 1994; Baan \& Haschick 1994).

Due to the fact that \Ic is the closest starburst galaxy to
our own Galaxy and has interesting properties at a
variety of wavelengths, we performed a deep pointing towards it
with the HRI detector (David et~al. 1995) on \ROSAT (Tr\" umper 1983).
Our main goals were to look for X-ray binaries, strong SNR, 
evidence for a hot interstellar medium and any
X-ray emission near the maser complexes. This is the first pointed 
X-ray observation of \Ic to our knowledge. 

Distance estimates to \Ic range from 0.7--3 Mpc, although a 
consensus is developing that the distance is $\approx 1$ Mpc 
(see Massey \& Armandroff 1995). We shall adopt this distance,
for which the HRI spatial resolution of $\approx 5$ arcsec 
corresponds to $\approx 24$~pc.  
The Galactic column density towards \Ic is high due
to its low Galactic latitude, and \Ic almost certainly has 
significant intrinsic absorption (see below). From the 
data of Stark et~al. (1992), we obtain a Galactic column 
density of $\approx 3.2\times 10^{21}$~cm$^{-2}$. 
This column blocks almost all photons with energies
$\simlt 0.4$ keV. 

NGC~147 and NGC~185 are a pair of dwarf elliptical galaxies that 
are about 100 kpc from M31. NGC~147 has a predominantly old 
stellar population ($\simgt 12$ Gyr) and 4 known globular clusters.
Its H~{\sc i} content is low. 
Most of the stars in NGC~185 are also old, but in addition it has
OB stars, dust clouds and a fairly 
massive amount of H~{\sc i} in its interstellar medium.
It probably contains a SNR (Gallagher, Hunter \& Mould 1984), and it 
has 6 known globular clusters. 
Unlike NGC~205, NGC~185 is too far from M31
to have had its star formation triggered by a recent interaction
with this galaxy. 
Hodge (1994) gives further details about these 
galaxies and references to the literature. Helfand (1984)
suggested that each of these galaxies might have $\sim 2$
X-ray binaries (see his table 1), so we performed \ROSAT HRI 
pointings towards them. We also wanted to look for any evidence
of a hot galactic wind, as discussed in section IVb of 
Ford, Jacoby \& Jenner (1977). For these galaxies we adopt a 
distance of 600 kpc and a Galactic column 
of $\approx 1.2\times 10^{21}$~cm$^{-2}$ (Stark et~al. 1992).

\section{Observations and data analysis} 

\subsection{\Ic}

\subsubsection{Observation and X-ray image details}

\Ic was observed with the \ROSAT HRI starting on 
1996 January 18 (RH600902: total raw exposure of 32.7 ks spread over 10.8
days). The \ROSAT observation was performed in the standard `wobble' mode,
and reduction and analysis of the resulting data were performed with the
Starlink {\sc asterix} X-ray data processing system. 

In Figure 1 we show contours of part of 
the full band ($\approx$0.1--2.5 keV) HRI image overlaid on 
the image of \Ic from the Palomar Schmidt $B_{\rm J}$ plate  
[see Minkowski \& Abell 1963 for more information on the
optical image]. We also show the positions of interesting
objects in \Ic identified at other wavelengths (see the figure caption).

We have checked the astrometry of the HRI image using a bright 
star that is also the brightest X-ray 
source in the field, and it appears 
to be nominal (see Figure 2). This 
star is identified as 
TV Cas (HD 1486; $V=7.3$; orbital period of 1.81 days), a 
well-studied, Algol-type, eclipsing 
binary (see Khalesseh \& Hill 1992 and references therein). 
In addition, we have also detected the W UMa-type binary
BH Cas ($V=12.3$; orbital period of $\sim 0.39$ days; see 
Metcalfe 1995 and references therein). The X-ray and 
SIMBAD database positions for BH Cas agree to 
within 3 arcsec, further confirming 
the HRI astrometry. These are the first X-ray detections of TV Cas 
and BH Cas to our knowledge, and we give their X-ray details in 
Appendix A.  

\subsubsection{\Ic X-1}

\noindent {\it Basic observed X-ray properties}
\vskip 0.1 in 

\noindent
The HRI image shows a clear detection of a pointlike 
source coincident with \Ic (hereafter referred to as \Ic X-1), and we have 
used the {\sc asterix} point source searching program {\sc pss} (Allan 1995) 
to quantify the source properties. The centroid position of X-1 is 
$\alpha_{2000}=$ 00$^{\rm h}$ 20$^{\rm m}$ 29.0$^{\rm s}$,
$\delta_{2000}=$ $59^\circ$16$^{\prime}$50.4$^{\prime\prime}$, 
with an error radius (statistical plus systematic) of 
approximately 5 arcsec. The source has $261\pm 18$ counts 
(after background subtraction), giving a mean 
HRI count rate of $7.9\times 10^{-3}$ count s$^{-1}$. The detection 
statistical significance is $>25\sigma$. It is possible to use
{\sc pimms} (Mukai 1995) and a spectral model to convert the
observed mean HRI count rate into a mean source flux. 
We adopt a power-law model with a photon index of $\Gamma=2$. 
The Galactic column density is $3.2\times 10^{21}$ cm$^{-2}$, and 
if we adopt this column density we obtain 
a power-law normalization of 
$2.4\times 10^{-4}$ photons cm$^{-2}$ s$^{-1}$ keV$^{-1}$ at 1 keV
and a 0.1--2.5~keV absorbed flux of 
$3.3\times 10^{-13}$ erg cm$^{-2}$ s$^{-1}$.
The 0.1--2.5~keV absorbed flux is not a strong function of the
adopted column density. For example, if we use a
much lower column of $5\times 10^{20}$ cm$^{-2}$,
then we obtain a power law normalization of
$1.1\times 10^{-4}$ photons cm$^{-2}$ s$^{-1}$ keV$^{-1}$ at 1 keV
and a 0.1--2.5 keV absorbed flux of 
$3.0\times 10^{-13}$ erg cm$^{-2}$ s$^{-1}$.
We estimate absorption-corrected fluxes and luminosities below,
after we have considered how likely it is that X-1 is indeed
associated with \Icc. 

\begin{figure}
\centerline{\psfig{figure=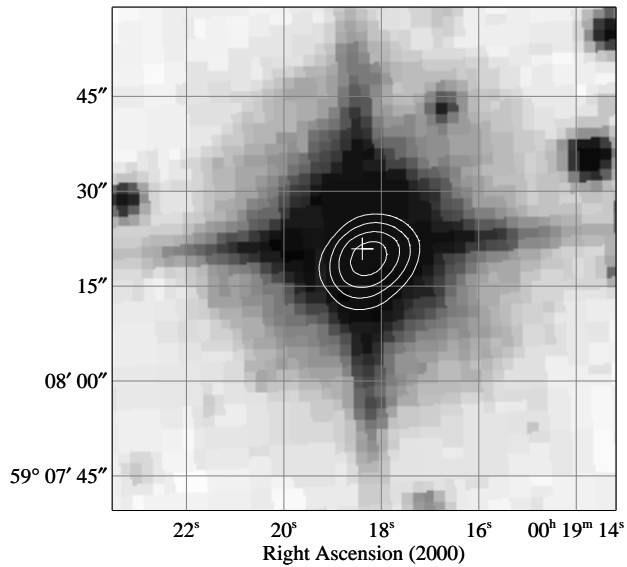,width=0.45\textwidth}}
\caption{Contours of part of our HRI image 
near TV Cas overlaid on the image from 
the Palomar Schmidt $B_{\rm J}$ plate. Contours are at 
58.1, 69.8, 81.4 and 93.0 per cent of the maximum pixel value (see 
Appendix A for absolute normalization). The cross shows the position 
of TV Cas from the SIMBAD database
($\alpha_{2000}=$ 00$^{\rm h}$ 19$^{\rm m}$ 18.4$^{\rm s}$,
$\delta_{2000}=$ $59^\circ$08$^{\prime}$20.9$^{\prime\prime}$). 
Note that the HRI contours are
well aligned with the star, confirming the HRI astrometry.}
\end{figure}

\begin{figure}
\centerline{\psfig{figure=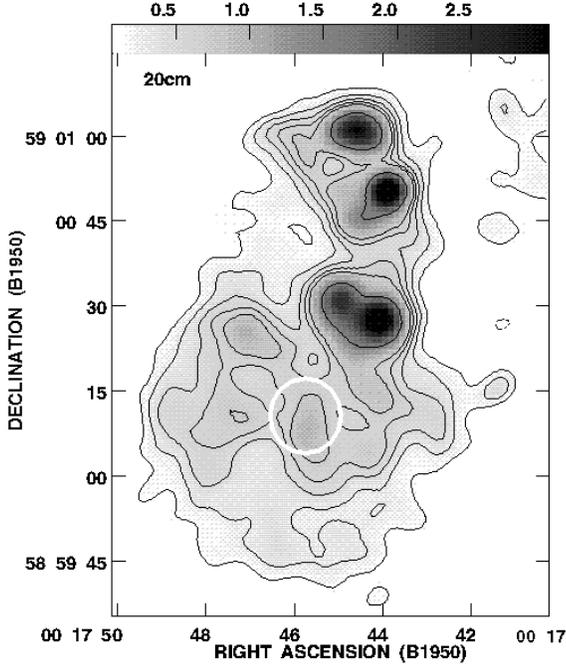,width=0.45\textwidth}}
\caption{Radio continuum image at 20 cm of the YS93 
non-thermal superbubble in \Ic (from YS93). 
The HRI error region for X-1 is marked as a thick, white
circle. The 1950 position for the centroid of  X-1 is 
$\alpha_{1950}=$ 00$^{\rm h}$ 17$^{\rm m}$ 45.9$^{\rm s}$,
$\delta_{1950}=$ $59^\circ$00$^{\prime}$11.9$^{\prime\prime}$.
The dark, black regions show sites of thermal radio 
emission associated with known H~{\sc ii} regions in
\Icc. The less intense emission to the south is the YS93
non-thermal superbubble. See YS93 for 
details of the radio observations. 
}
\end{figure}

We have searched the literature for sources at other wavelengths 
that are coincident with the detected X-ray source, and we find that
the X-ray source lies in the middle of a hive of activity in \Icc. 
YS93 quote an approximate central 
position for their non-thermal radio superbubble of 
$\alpha_{2000}=$ 00$^{\rm h}$ 20$^{\rm m}$ 28$^{\rm s}$,
$\delta_{2000}=$ $59^\circ$16$^{\prime}$49$^{\prime\prime}$, 
and X-1 is within 7.5 arcsec of
this position (see Figure 3). It is 
within 4.1 arcsec of the strongest
`SNR candidate' listed in table 3 of YS93. 
The H~{\sc ii} region \# 113 from 
Hodge \& Lee (1990) has a central position of  
$\alpha_{2000}=$ 00$^{\rm h}$ 20$^{\rm m}$ 27.7$^{\rm s}$,
$\delta_{2000}=$ $59^\circ$16$^{\prime}$50$^{\prime\prime}$
(after making the correction described in Ohta et~al. 1992), 
which is about 10 arcsec from the HRI centroid position. 
This H~{\sc ii} region is extended and the catalogued 
H~{\sc ii} regions in this area are underlain by widespread
diffuse emission. 
In addition, the peculiar 
star WR17 from Massey \& Armandroff (1995) is located at
$\alpha_{2000}=$ 00$^{\rm h}$ 20$^{\rm m}$ 29$^{\rm s}$,
$\delta_{2000}=$ $59^\circ$16$^{\prime}$52$^{\prime\prime}$,
which is within 2 arcsec of the X-ray centroid.
Massey \& Armandroff (1995) give an optical spectrum of 
this star and classify it as a WN type Wolf-Rayet star
based on the detection of a broad feature which they 
identify as He {\sc ii} 4686~\AA. 
They find a $V$ magnitude of 21.76 and
$E(B-V)=0.77$. Using the 3.5-m telescope of the 
Apache Point Observatory, we have performed
independent optical spectroscopy which confirms that 
WR17 is an emission-line star of some type. 
The H$\alpha$ and [N~{\sc ii}] emission lines of WR17 
are systematically shifted by $75\pm 9$ km s$^{-1}$
to the red compared to those of the aforementioned diffuse 
emission. This rules out the possibility that WR17 is a normal 
supergiant star with emission lines superimposed on it from the 
diffuse emission. The measured velocity for WR17 is
$-338$ km s$^{-1}$, and this agrees better with that
of the surrounding H~{\sc i} than that of the
H~{\sc ii}.
WR18 is also located within 6 arcsec of the HRI centroid,
although Massey \& Armandroff (1995) state that this is
probably not a Wolf-Rayet star. Our independent spectroscopy
suggests that WR18 may be an ordinary supergiant star with
emission lines superimposed on it from the H~{\sc ii} region.

\vskip 0.1 in
\noindent {\it Confusing foreground and background sources}
\vskip 0.1 in 

\noindent
\Ic is at low Galactic latitude so confusing foreground sources 
are a natural concern. Based on results from the \ROSAT Galactic
Plane survey (see section 9.1 of Motch et~al. 1997), we expect
about 0.9 sources per square degree with count rates as large as 
or larger than that of X-1. \Ic has an angular size of
about $6.3\times 5.1$ arcmin$^2$, so we estimate that there 
is less than about a 1 per cent chance of having a 
confusing foreground source this bright in front of \Icc.
Furthermore, the HRI source is found at a position where 
one might expect X-ray activity in \Ic based on its emission 
in other wavebands. For example, the X-ray source is almost
exactly in the centre of the YS93 non-thermal superbubble 
(see YS93 for good arguments, but not rigorous proof, that the 
superbubble itself is in \Icc) and near a region of recent 
star formation. The superbubble has an angular size of
about $0.7\times 0.7$ arcmin$^2$, so there is less than
about a 0.02 per cent chance of having a confusing foreground 
source as bright as X-1 in front of it. We can further
argue against some types of possible foreground sources based 
on examination of optical images. For example, we can use
the nomograph shown in figure~1 of Maccacaro et~al. (1988)
which relates the expected 0.3--3.5~keV X-ray flux to the
visual magnitude for various types of astronomical objects. 
If we use the power-law model with a column of
$5\times 10^{20}$ cm$^{-2}$ (see above), we predict a
0.3--3.5~keV flux of 
$\approx 3.5\times 10^{-13}$ erg cm$^{-2}$ s$^{-1}$.  
There are no optical stars in the X-ray error circle with
$V$ magnitudes brighter than 20.8
(based on analysis of a $V$ image taken with
the 6-m telescope of the Russian Academy of Sciences), and 
thus from figure~1 of Maccacaro et~al. (1988) we see that the
combined X-ray flux and $V$ magnitude limit do not agree 
well with those expected from a normal foreground star
(this is true even for 0.3--3.5~keV fluxes as low as 
$\approx 1.9\times 10^{-13}$ erg cm$^{-2}$ s$^{-1}$).
Motch et~al. (1997) state that about 85 per cent of their 
sources from the \ROSAT Galactic Plane survey are stars
with X-ray emitting coronae, so we have ruled out the
most probable type of confusing foreground source. 
In summary, although we cannot formally prove that a 
foreground contaminating source is impossible, we consider one
to be quite unlikely.

Confusing background sources (e.g. Seyfert galaxies, 
quasars and clusters of galaxies)
are also unlikely. Examination of figure~18 of 
Motch et~al. (1997) shows that the probability of
having an extragalactic source as bright as X-1
behind the superbubble is $\simlt 1\times 10^{-4}$.
This is a conservative probability estimate since any
extragalactic source would also suffer from X-ray
absorption by matter within the plane of \Icc.
Some types of possible extragalactic confusing 
sources (e.g. clusters of galaxies) can also be
ruled out based on the X-ray variability 
described below.

\begin{figure}
\centerline{\psfig{figure=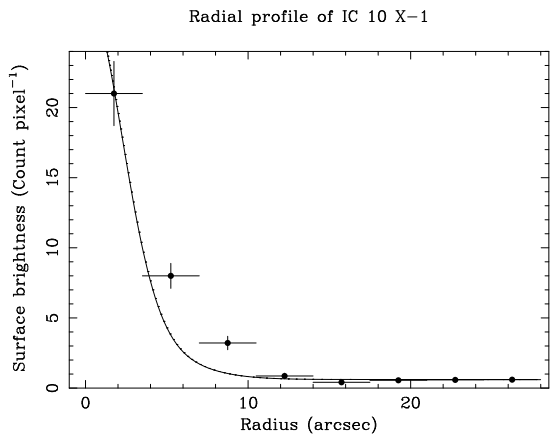,width=0.45\textwidth,angle=270}}
\caption{The data points show the
HRI radial profile for \Ic X-1. The background level
of $\approx 0.6$ count pixel$^{-1}$ is reached by a 
radius of $\approx 14$ arcsec. The solid line shows the
nominal HRI point spread function from 
section~2.2.3 of David et~al. (1995). This
point spread function has been normalized to
the first data point for \Ic X-1.
Note that any X-ray extent is much
less than the extent of the non-thermal 
radio superbubble. The small extent suggested
by the data may well be entirely due to aspect 
solution errors (see the text for details).
}
\end{figure}

\vskip 0.1 in 
\noindent {\it Luminosity, spatial extent and variability}
\vskip 0.1 in 

\noindent
Given that X-1 is probably located in the plane of \Icc, we expect
that it will have some absorption over the Galactic column of
$3.2\times 10^{21}$ cm$^{-2}$. We expect an intrinsic
column in \Ic of $\sim 2\times 10^{21}$ cm$^{-2}$ based on figure 5 
of YS93. We thus adopt a total column of $5.2\times 10^{21}$ cm$^{-2}$.
Using this column and a power-law model 
with a photon index of $\Gamma=2$,
we obtain a power-law normalization of 
$3.5\times 10^{-4}$ photons cm$^{-2}$ s$^{-1}$ keV$^{-1}$
at 1 keV. The 0.1--2.5 keV absorbed flux for this model
is $3.7\times 10^{-13}$ erg cm$^{-2}$ s$^{-1}$,  
and the 0.1--2.5 keV absorption-corrected flux is 
$1.8\times 10^{-12}$ erg cm$^{-2}$ s$^{-1}$.
While we note that the flux correction factor 
used above is large (about 4.9), the model we have used 
seems physically appropriate given the probable nature 
of X-1 (see below), and the presence of a large
column between the plane of \Ic and Earth is without 
question. At a distance of 1 Mpc the absorption-corrected 
flux corresponds to a mean isotropic 
luminosity of $2.0\times 10^{38}$ erg s$^{-1}$.
If, for comparison, we only use the Galactic column and neglect the
probable absorption in \Ic then we obtain a  
0.1--2.5 keV absorption-corrected flux of 
$1.2\times 10^{-12}$ erg cm$^{-2}$ s$^{-1}$
and an isotropic luminosity of 
$1.4\times 10^{38}$ erg s$^{-1}$ (the power-law
normalization and absorbed flux for this model
are given above). 

Inspection of the X-ray morphology of X-1 does not show any 
strong evidence for deviations from a circular shape.
As the YS93 superbubble has an extent of $\approx 45$ 
arcsec, we have examined the spatial extent of 
\Ic X-1 (see Figure~4). \Ic X-1 is much less extended
than the radio emission from the superbubble. While the radial
profile shown in Figure~4 is not formally consistent with
the nominal point spread functions described 
in David et~al. (1995), it is known that aspect solution 
errors can lead to a significant broadening of the profiles of
point sources in some HRI observations (see the 
discussion in Morse 1994). We do
not have enough counts to perform aspect error correction using
the Morse code {\sc hriaspcor} in {\sc ftools}, and TV Cas is
too far off axis to be a useful comparison point source.  
The current data do not give strong evidence for spatial
extent over that from a point source which suffers from 
aspect solution errors. Furthermore, the variability
discussed below suggests that most of
the X-ray flux originates from a compact region with an
extent of less than $\sim 1$ light day.

Count rates from \ROSAT should be averaged over an integer multiple of 
the 400-s wobble period when used to search for rapid variability of 
cosmic X-ray sources (see Brinkmann et~al. 1994; although note that the 
effects of the wobble are less serious for the HRI than the PSPC).
We have used a source cell with a radius of 20 arcsec to extract 
a light curve of X-1. We use 400-s bins and neglect 
partially filled bins. In Figure 5 we show part of this light curve 
where there is evidence for X-ray variability. The count rate from
\Ic X-1 appears to have dropped by over a factor of 2 within a day.  
We have calculated the Poisson probability that the variability
is merely due to photon statistics, and it is less than 
$1\times 10^{-8}$. Even taking into account issues such as
those described in Press \& Schechter (1974), the
variability appears to be of high statistical significance. 
We have used a double panel format 
in Figure 5 for clarity because our observations 
are spread over 10.8 days, and there are no data in between the two 
panels shown in Figure 5. We have verified the variability shown in
Figure 5 using two independent \ROSAT analysis software systems. 
The mean background count rate in the source 
cell, computed using a large, nearby, circular, background cell, is 
$\approx 1.5\times 10^{-3}$ count s$^{-1}$ for the left panel and
$\approx 1.3\times 10^{-3}$ count s$^{-1}$ for the right panel.
We have verified that there is no positional error 
(e.g. a shift or rotation) that could be
moving the source out of our source cell during the times shown
in our right hand panel (see Figure 6). If we bin an image using only the data from 
the left hand panel, we can clearly see X-1 over the background
level. However, if we bin an image using only the data from
the right hand panel, we cannot see X-1 despite the 
fact that the right hand panel has twice as much exposure and a  
slightly lower background. There is evidence for variability
similar to that shown in Figure 5 throughout our observation. 
The highest count rate observed is about twice the mean
count rate, which corresponds to a 0.1--2.5 keV isotropic 
luminosity of $\approx 4\times 10^{38}$ erg s$^{-1}$.   

As discussed below, it appears likely that X-1 is an X-ray
binary system in \Icc. Unfortunately, our data do not appear
to provide strong constraints on the binary orbital period
via periodic X-ray flux variations (e.g. eclipses). This is due
partially to the low count rate, but also to the fact that our
32.7~ks of exposure is broken into numerous, small, separated
blocks with durations between 1200--3000~s. 

\subsubsection{Other X-ray emission in \Ic }

We have used {\sc pss} and the HRI point spread function to search 
our HRI field for any other X-ray point sources coincident with the
extent of \Icc. We do not detect any point X-ray sources stronger 
than $4.5\sigma$ within the extent shown in figure 5 of YS93. 
We do not detect any significant point sources near the maser 
sites, and we do not detect the radio continuum source described
in section 3.3 of Argon et~al. (1994). The \ROSAT Standard
Analysis Software System (SASS) reports a possible diffuse X-ray 
source at
$\alpha_{2000}=$ 00$^{\rm h}$ 20$^{\rm m}$ 18.6$^{\rm s}$,
$\delta_{2000}=$ $59^\circ$18$^{\prime}$50$^{\prime\prime}$
with a positional error of 20 arcsec. 
Careful examination of the X-ray image and the SASS output
shows that this source is not of high statistical
significance in the current data (e.g. there are background 
fluctuations at almost comparable levels). However, the fact
that the SASS X-ray source is within 12.5 arcsec of a bright 
radio source discussed in section 3.4 of Klein \& Gr\" ave (1986) 
suggests that it may well be real (note that some of the source 
labeling in section 3.4 of Klein \& Gr\" ave 1986 is 
incorrect). The SASS X-ray source is also quite close
to two of the sources listed in table 1 of YS93.  
This is the location of an H~{\sc i} hole in \Ic 
which may have been created by supernova events. The
`best guess' HRI count rate for this source is 
$\sim 8\times 10^{-4}$ count s$^{-1}$. Further X-ray 
observations are needed to study this emission. 

We have used {\sc pss} and the HRI point spread function to 
place upper limits on point sources at several representative 
positions in \Ic that lie $>14$ arcsec from 
\Ic X-1. The typical $2\sigma$ upper limit is about 
$3.0\times 10^{-4}$ count s$^{-1}$. Using a $\Gamma=2$ power
law with a column of $4.2\times 10^{21}$ cm$^{-2}$
(motivated by the Galactic column and figure 5 of YS93),
this corresponds to an absorbed 0.1--2.5 keV flux upper limit 
of about $1.3\times 10^{-14}$ erg cm$^{-2}$ s$^{-1}$ 
and an absorption-corrected 0.1--2.5 keV flux upper limit of about    
$5.7\times 10^{-14}$ erg cm$^{-2}$ s$^{-1}$. 
The isotropic luminosity upper limit is about
$6.4\times 10^{36}$ erg s$^{-1}$. 
This upper limit is appropriate for X-ray sources that
are extended on scales smaller than or comparable to
that of the HRI spatial resolution (about 5 arcsec
which corresponds to about 24 pc at the distance of
\Icc). Thus, this limit can be applied to supernova
remnants in the free-expansion or Sedov-Taylor phases
(with ages $\simlt 5000$ yr), although many supernova 
remnants have luminosities lower than our upper limit 
(see table~3.2 of Charles \& Seward 1995).
Of course, it is possible that there 
are even more highly obscured X-ray sources in \Ic 
(i.e. sources that suffer even stronger intrinsic absorption 
in \Ic itself) with larger isotropic X-ray luminosities than 
the limit given above. A hard X-ray observation of \Ic could 
search for such sources. 

In order to search for diffuse emission from \Ic
(e.g. from a hot interstellar medium), we have carefully
examined moderately smoothed and heavily smoothed
HRI images. We have used both full-band HRI images
as well as images made using only HRI channels 3--8
(this can help to maximize the signal-to-noise level;
see section 2.4 of David et~al. 1995). There does not
appear to be any highly significant diffuse X-ray emission
coincident with the optical or radio extents of
\Icc. 
To obtain a characteristic constraint on
the amount of large-scale diffuse emission, we consider
a circular `source' aperture centred at
$\alpha_{2000}=$ 00$^{\rm h}$ 20$^{\rm m}$ 16.2$^{\rm s}$,
$\delta_{2000}=$ $59^\circ$18$^{\prime}$23$^{\prime\prime}$
with a radius of 100 arcsec. This aperture includes
many of the H~{\sc ii} regions of \Icc, and it does
not include \Ic X-1. The mean count rate density within
this aperture is
$(4.72\pm 0.14)\times 10^{-3}$ count s$^{-1}$ arcmin$^{-2}$.
For comparison, a circular `background' aperture
that is located outside the extent of \Ic (at
$\alpha_{2000}=$ 00$^{\rm h}$ 20$^{\rm m}$ 55.6$^{\rm s}$,
$\delta_{2000}=$ $59^\circ$23$^{\prime}$57$^{\prime\prime}$
and with a radius of 112 arcsec)
has a mean count rate density of
$(4.54\pm 0.11)\times 10^{-3}$ count s$^{-1}$ arcmin$^{-2}$
(this mean count rate density may be compared with the HRI
background values listed in table~5 of David et~al. 1995). 
Thus, after background subtraction, our source aperture
has a residual mean count rate density of 
$(1.8\pm 1.8)\times 10^{-4}$ count s$^{-1}$ arcmin$^{-2}$
(i.e. it is consistent with zero).
Adopting the same spectral model as used above for
point-source upper limits, we find a mean (absorbed) 
surface brightness of
$(7.8\pm 7.8)\times 10^{-15}$ erg cm$^{-2}$ s$^{-1}$ arcmin$^{-2}$.
Of course, this value is an average
constraint on large-scale diffuse emission, and it 
cannot be used on a point-by-point basis. 

\subsubsection{The possible all-sky survey detection of \Ic }

\Ic is listed (without discussion) in Boller et~al. (1992) 
as having been detected during the \ROSAT all-sky 
survey (RASS). Revision of the all-sky survey processing
has led to an improved position for the RASS source of 
$\alpha_{2000}=$ 00$^{\rm h}$ 20$^{\rm m}$ 21$^{\rm s}$,
$\delta_{2000}=$ $59^\circ$16$^{\prime}$18$^{\prime\prime}$
(Th. Boller, private communication),
and we show this position in Figure 1. The \ROSAT Positional 
Sensitive Proportional Counter (PSPC; Pfeffermann et~al. 1987)  
registered $0.023\pm 0.008$ count s$^{-1}$ from 
the claimed all-sky survey
source (Th. Boller, private communication), which 
corresponds to an equivalent HRI count rate of about
$8\times 10^{-3}$ count s$^{-1}$. The all-sky survey exposure
was short (494 s; Th. Boller, private communication) and 
places no useful spectral constraints on the PSPC 
source. Only about 13 PSPC photons in total 
(including background) were collected.
The revised RASS position is 
1.15 arcmin from our HRI position. The 
equivalent HRI count rate for the RASS source and 
our mean pointed HRI count rate for X-1 
are similar, but the positional offset is 
somewhat larger than would be expected from the 
typical $\approx 5$ arcsec HRI error circle radius and the
typical $\approx 40$ arcsec PSPC error circle radius.
There are no HRI sources formally consistent with the 
RASS position, despite the fact that the HRI image
probes much deeper than the RASS. Our position for X-1 is 
almost certainly not in error (see Figure 2
and the associated discussion), and the RASS data do
not place a strong limit on the brightness of X-1
during the all-sky survey (X-1 lies between the 58
and 69 per cent contour levels for the RASS source).

\begin{figure}
\centerline{\psfig{figure=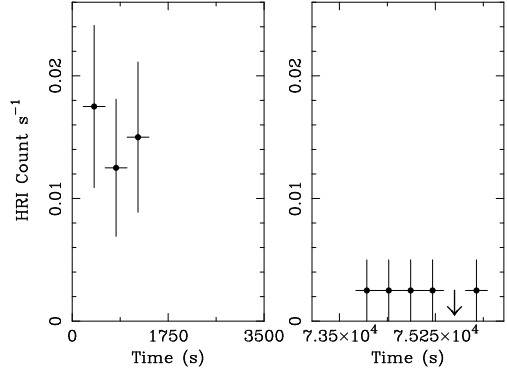,width=0.45\textwidth,angle=270}}
\caption{HRI light curve of \Ic X-1 from part of our observation.
Times are measured from 1996 January 18 16:13:20 UT.
The light curve bin size is 400 s, and the two
abscissae span the same length of time. 
}
\end{figure}

\subsection{NGC~147 and NGC~185}

NGC~147 was observed with the HRI starting on
1995 January 19 (RH400744: total raw exposure of 14.8 ks 
spread over 63.6 ks), and NGC~185 was also observed starting
on 1995 January 19 (RH400743: total raw exposure of 21.0 ks
spread over 91.1 ks). Each observation had its pointing 
position coincident with the optical centre of the 
appropriate galaxy. 

\begin{table}
\caption{X-ray sources near NGC~147 (the first three sources) 
and NGC~185 (the last two sources) in the HRI fields. Note
that many of these sources are probably not related to NGC~147 or
NGC~185 (see the text). The last column gives a brief description of
the optical morphology based on inspection of Palomar Schmidt plates,
Palomar 200-inch plates and Lick 3-m plates.}
\begin{center}
\begin{tabular}{cllll}
\hline
Source  &                  &                  & HRI            & Optical \\
Number  & $\alpha_{2000}$  & $\delta_{2000}$  & count s$^{-1}$ & Morph.  \\
\hline
1 & 00 32 10.6 & 48 33 40 & $1.2\times 10^{-3}$  & Diffuse   \\
2 & 00 32 28.3 & 48 31 56 & $2.5\times 10^{-3}$  & Diffuse \\
3 & 00 34 02.1 & 48 23 38 & $6.5\times 10^{-4}$  & Point source \\
4 & 00 38 21.9 & 48 24 09 & $8.0\times 10^{-4}$  & See the text \\
5 & 00 39 02.3 & 48 24 15 & $1.5\times 10^{-3}$  & Point source \\
\hline
\end{tabular}
\end{center}
\end{table}

\begin{figure*}
\hbox{
\psfig{figure=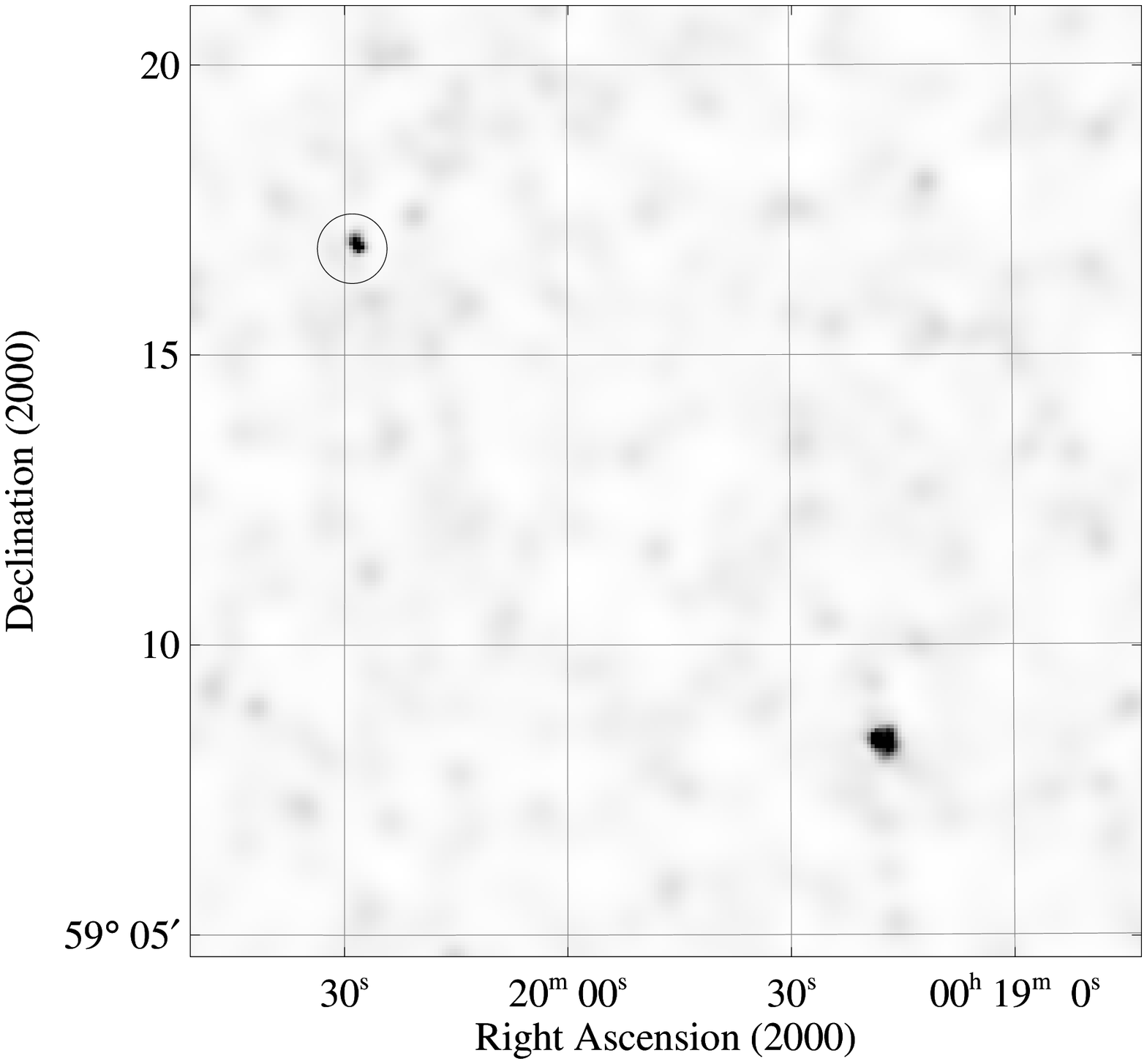,width=0.5\textwidth}
\psfig{figure=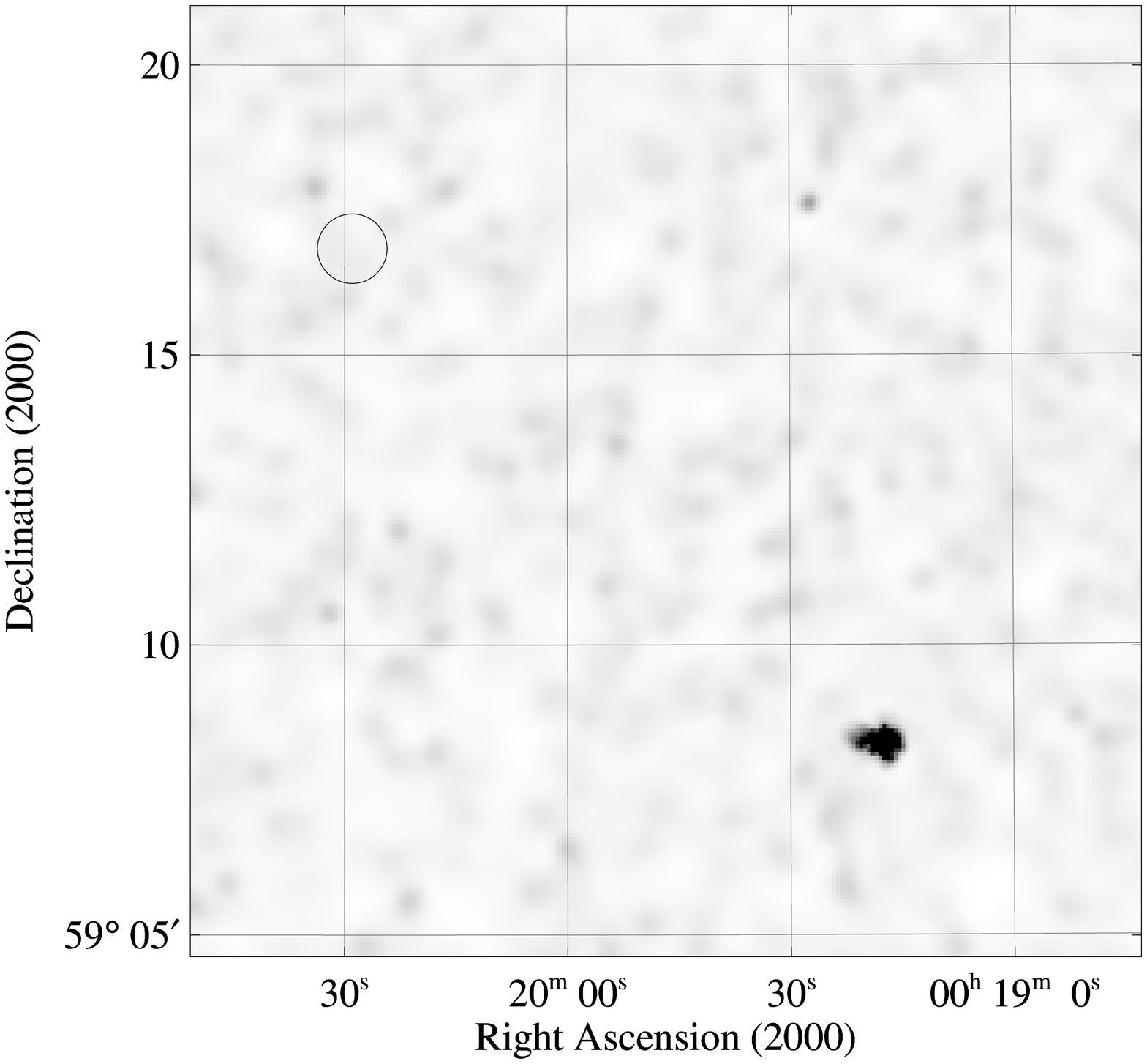,width=0.5\textwidth}
}
\caption{Smoothed HRI images illustrating the observed
variability of X-1. X-1 is the circled source
(the circle radius is arbitrary), and
TV Cas is the source to the lower right.  
The left hand panel corresponds to the
left hand panel of Figure 5, and the right hand panel 
corresponds to the right hand panel of Figure 5.
The right hand panel has a longer integration time
than the left hand panel, and the two images have the 
same (linear) scaling. Note that X-1 clearly varies
in HRI count rate.}
\end{figure*}

We have used {\sc pss} and the HRI point spread function to 
search for X-ray point sources coincident with these galaxies. 
We do not detect any X-ray sources coincident with the
cores of these galaxies, and we do not detect any X-ray 
sources coincident with their known globular clusters either
(we use the globular cluster positions from table 9 of
Ford et~al. 1977). As with \Icc, we have scrutinized
smoothed X-ray images of each of these galaxies and do not
see any evidence for diffuse emission. We have placed
upper limits on point sources at several representative
positions in NGC~147, and the typical 2$\sigma$ upper
limit is about $6.0\times 10^{-4}$ count s$^{-1}$. 
Using a $\Gamma=2$ power law with the Galactic 
column ($1.2\times 10^{21}$~cm$^{-2}$), this corresponds 
to an absorbed 0.1--2.5 keV flux upper limit of about
$2.3\times 10^{-14}$ erg cm$^{-2}$ s$^{-1}$
and an absorption-corrected 0.1--2.5 keV flux upper limit of about
$5.7\times 10^{-14}$ erg cm$^{-2}$ s$^{-1}$. The isotropic 
luminosity upper limit is about $2.3\times 10^{36}$ erg s$^{-1}$. 
We have performed the same procedure for NGC~185, and the
corresponding upper limits are 
$4.6\times 10^{-4}$ count s$^{-1}$ (count rate),
$1.7\times 10^{-14}$ erg cm$^{-2}$ s$^{-1}$ (absorbed flux),
$4.3\times 10^{-14}$ erg cm$^{-2}$ s$^{-1}$ (absorption-corrected flux) and
$1.7\times 10^{36}$ erg s$^{-1}$ (isotropic luminosity). 

In Table 1 we list all X-ray sources detected within 12 arcmin of
the centres of NGC~147 and NGC~185. All of the sources in this
table lie outside the central optical extents of NGC~147 and
NGC~185, and all but source 5 lie outside the regions 
previously searched for globular clusters 
(see Baade 1951; Hodge 1974; Hodge 1976; Ford et~al. 1977).
However, these galaxies both are optically very large 
at faint surface brightness levels. Most of the 
X-ray sources in Table~1 are 
probably unrelated foreground or background
sources, but a couple of them may be associated with these 
galaxies or currently unknown 
globular clusters. None of the sources 
in Table~1 is currently listed in the 
NASA Extragalactic Database (NED) or the SIMBAD database. 
We give optical morphologies of possible counterparts
to the X-ray sources in Table~1.
Source 4 is near the center of a ring of stellar images and
there appears to be a very faint object about 2 arcsec southwest 
of the X-ray position.
Source 5 is almost certainly not a globular cluster, as it is
not resolved into stars on red Palomar plates. It appears to
be a foreground star.

\section{Discussion and summary} 

X-1 appears to be located in \Ic near the centre 
of the YS93 non-thermal radio 
superbubble, and it has a powerful mean 
0.1--2.5 keV isotropic luminosity of 
$\approx 2\times 10^{38}$ erg s$^{-1}$. Single 
(nondegenerate) stars do not produce this much X-ray 
emission (e.g. Rosner, Golub \& Vaiana 1985; 
Pollock 1987; Kashyap et~al. 1992; and 
references therein), and 
the X-ray luminosities of cataclysmic variables also 
fall short by several orders of magnitude
(e.g. Eracleous, Halpern \& Patterson 1991). The observed 
variability argues against most of the emission coming from a 
cluster of stars or a supernova remnant. Thus it
appears most likely that we have discovered an accreting
neutron star or black hole in a binary system. The 
highest 0.1--2.5 keV luminosity we observe from \Ic X-1 
is $\approx 4\times 10^{38}$ erg s$^{-1}$,
and the total X-ray luminosity is likely to be  
$\simgt 6\times 10^{38}$ erg s$^{-1}$ if the source is indeed
an accreting neutron star or black hole (see 
White, Nagase \& Parmar 1995 for a discussion of the X-ray
spectral energy distributions of X-ray binaries). 
This is the Eddington limit 
luminosity for a $\approx 5$ M$_{\odot}$ object which is 
above the maximum mass for a neutron star. While 
this may suggest the compact object
is a black hole, we cannot rule out a neutron star which
emits anisotropically or at a super-Eddington rate. This 
is especially true in light of the low heavy element 
abundances in \Ic (e.g. Lequeux et~al. 1979) and the apparent
X-ray luminosity/abundance relation for accreting compact
objects (Clark et~al. 1978; section 2.5 of 
van Paradijs \& McClintock 1995). 

As a peculiar emission-line star (WR17) has been 
located in the tiny HRI error circle by 
Massey \& Armandroff (1995; see above), it is worth 
investigating whether this could be the binary
companion star to the accreting compact object. 
If, as claimed by Massey \& Armandroff (1995),
this star is a WN type Wolf-Rayet star, then \Ic X-1 would be a
more X-ray luminous version of the possible Wolf-Rayet 
X-ray binaries reported near the centre of 
30 Doradus (Wang 1995; also see van Kerkwijk et~al. 1996
and references therein
regarding the possible Wolf-Rayet nature of Cyg X-3). The
system would have originally been a binary with two massive stars, 
and a supernova in this system would probably have contributed 
to the creation of the YS93 superbubble (such a system might
plausibly remain bound after a supernova; see 
Brandt \& Podsiadlowski 1995).  
Of course, the identification of
X-1 with WR17 has not yet been proven and X-1 could also be
a low mass X-ray binary in which the optical companion is below
the current threshold of detectability 
(cf. Stocke, Wurtz \& K\"uhr 1991; but note that
Petre 1993 suggests that low mass X-ray binaries
are rare in star forming regions).  

Finally, the presence of a powerful X-ray source in the
centre of an unusually large bubble of radio emission
invites a brief comparison with SS433 and W50. 
YS93 compared the radio surface brightness ($\Sigma$)
and diameter (d) of their superbubble to 
the so-called `$\Sigma$-d relationship' for
SNR in the Magellanic Clouds. 
While the physical meaning, if any, of the 
$\Sigma$-d relationship is somewhat 
obscure (especially when transported wholesale
from the Magellanic Clouds to \Icc), the 
YS93 superbubble does appear to
have a much larger diameter than expected given 
its radio surface brightness. YS93 
proposed a multiple SNR 
model to explain this fact. 
While the YS93 model appears entirely 
plausible, W50 also 
lies off the $\Sigma$-d relation (Margon 1984; its largest
dimension is about 65 pc). The growth of the YS93 superbubble
could have been influenced by \Ic X-1
(see section 5 of Margon 1984). \Ic X-1 does 
not appear to have redshifted/blueshifted emission lines in
the spectrum of Massey \& Armandroff (1995) and there is
no strong radio point source at its position, so if there
were SS433-like activity in the past it appears 
to have ended. 

\ASCA or {\it SAX\/} observations 
of \Ic would be very useful as they would
(1) probably obtain a significantly higher count rate from \Ic X-1
due to the heavy column which strongly affects the HRI band,
(2) allow a better variability study including searches for
eclipses from a high mass companion and X-ray pulsations,
(3) allow a measurement of the spectral shape of \Ic X-1 and
(4) search for additional variable and highly absorbed X-ray
sources in \Icc. They would also allow a much more precise
determination of the luminosity of \Ic X-1 since the 
column correction factor would not be so large.  
We are attempting to obtain such observations.
We are also trying to search for evidence of a binary star 
(especially WR17) within the X-ray error circle and 
obtain higher quality spectra to further examine the Wolf-Rayet 
classification of WR17 by Massey \& Armandroff (1995). 
Our current spectra do not allow definitive discussions of
either of these issues. 

If there are any X-ray binaries in the centres of 
NGC~147 or NGC~185, they were
quite faint at the times of these observations (see 
Section 2.2 for upper limits). We do not detect the putative SNR 
of Gallagher et~al. (1984), although our data do not
rule out the possibility that this object is a SNR.  

\section*{Acknowledgments}

We gratefully acknowledge financial support from the 
Smithsonian Institution and the United States 
National Science Foundation (WNB) 
and the Royal Society (ACF). 
We thank H. Ebeling for the use of his {\sc imcont} software,
and we thank D. Zucker for reducing the optical spectra of 
stars in the region of \Ic X-1.
We thank 
Th. Boller, 
D. Harris, 
L. Ho,
P. Massey, 
J. McClintock,
P. O'Brien, 
R. Petre,
Ph. Podsiadlowski and 
H. Yang for helpful discussions.  
The Palomar Schmidt plate image was obtained via the 
Digitized Sky Survey which was produced at the 
Space Telescope Science Institute under United States Government 
grant NAG W-2166. The Palomar Observatory Sky Survey was funded
by the National Geographic Society.

\appendix 
\section{Basic X-ray data for TV Cas and BH Cas}

In this appendix we give the first X-ray data on the 
serendipitously detected binaries TV Cas and BH Cas (see 
section 2.1.1). This will allow their inclusion in future
systematic studies of the X-ray emission from Algol and 
W UMa-type binaries (e.g. Singh, Drake \& White 1996;
McGale, Pye \& Hodgkin 1996; and references therein).  

\subsection{TV Cas: Algol-type binary}

TV Cas is offset from the centre of the HRI field of view by about
12 arcmin, and we detect about 1080 counts from it after background
subtraction (the detection 
significance is $>50\sigma$). The mean count rate 
after correction for vignetting is $3.6\times 10^{-2}$ count s$^{-1}$, 
and there is evidence for count rate variability by a factor of 
about 2. The low count rate and large data gaps prevent a detailed 
X-ray variability analysis. Based on the count rate conversion
factor in Singh et~al. (1996) and the relative responses of the
\ROSAT PSPC and HRI, we estimate that 
1 HRI count s$^{-1}$ corresponds to about
$3.0\times 10^{-11}$ erg cm$^{-2}$ s$^{-1}$. Using a distance
of 275 pc (from table 7 of Khalesseh \& Hill 1992), we 
calculate a 0.1--2.5 keV luminosity of about
$9\times 10^{30}$ erg s$^{-1}$, which suggests fairly
typical coronal activity for an Algol-type binary. 
  
\subsection{BH Cas: W UMa-type binary}

BH Cas is offset from the centre of the HRI field of view by about
11 arcmin, and we detect about 42 counts from it after background
subtraction (the detection significance is $5.4\sigma$). The count rate 
after correction for vignetting is $1.4\times 10^{-3}$ count s$^{-1}$,
and we are not able to probe for variability due to the low count rate. 
We adopt the same count rate to flux conversion factor as above and thus
derive a 0.1--2.5 keV flux of $4.2\times 10^{-14}$ erg cm$^{-2}$ s$^{-1}$. 
The distance to BH Cas is not well known, but an upper limit on the
distance is $530\pm 70$ pc (T. Metcalfe, private communication). Thus the 
X-ray luminosity appears to be less than about 
$1.3\times 10^{30}$ erg s$^{-1}$. The observed X-ray flux does
not allow a strong constraint on the distance via the 
implied luminosity (compare the range of W UMa X-ray 
luminosities seen in McGale et~al. 1996).  

\bsp

\end{document}